\documentclass[pr,cha,onecolumn,superscriptaddress,preprintnumbers,showpacs, amsmath,amssymb]{revtex4-1}
\usepackage{bm}
\usepackage[colorlinks=true,linkcolor=blue,citecolor=blue]{hyperref}
\usepackage{amsmath}
\usepackage{amssymb}
\usepackage{amsthm}
\usepackage{amsfonts}
\usepackage{enumerate}
\usepackage{latexsym}
\usepackage{ifpdf}
\usepackage{graphicx}
\usepackage{makeidx}
\expandafter\ifx\csname package@font\endcsname\relax\else
\expandafter\expandafter
\expandafter\usepackage
\expandafter\expandafter
\expandafter{\csname package@font\endcsname}
\fi
\usepackage{color}
\usepackage{times}

\begin{document}

\title{Transformation of hexagonal Lu to cubic LuH$_{2+x}$ single-crystalline films}

\author{Peiyi Li}
\thanks{These authors contributed equally to this work}
\affiliation{School of Materials Science and Chemical Engineering, Ningbo University, Ningbo 315211, PR China }
\affiliation{Ningbo Institute of Materials Technology and Engineering, Chinese Academy of Sciences, Ningbo 315201, China}
\author{Jiachang Bi}
\thanks{These authors contributed equally to this work}
\affiliation{Ningbo Institute of Materials Technology and Engineering, Chinese Academy of Sciences, Ningbo 315201, China}
\author{Shunda Zhang}
\affiliation{Ningbo Institute of Materials Technology and Engineering, Chinese Academy of Sciences, Ningbo 315201, China}
\author{Rui Cai}
\affiliation{Ningbo Institute of Materials Technology and Engineering, Chinese Academy of Sciences, Ningbo 315201, China}
\author{Guanhua Su}
\affiliation{Ningbo Institute of Materials Technology and Engineering, Chinese Academy of Sciences, Ningbo 315201, China}
\author{Fugang Qi}
\affiliation{Ningbo Institute of Materials Technology and Engineering, Chinese Academy of Sciences, Ningbo 315201, China}
\author{Ruyi Zhang}
\affiliation{Ningbo Institute of Materials Technology and Engineering, Chinese Academy of Sciences, Ningbo 315201, China}
\author{Zhiyang Wei}
\affiliation{Ningbo Institute of Materials Technology and Engineering, Chinese Academy of Sciences, Ningbo 315201, China}
\author{Yanwei Cao}
\email{ywcao@nimte.ac.cn}
\affiliation{Ningbo Institute of Materials Technology and Engineering, Chinese Academy of Sciences, Ningbo 315201, China}
\affiliation{Center of Materials Science and Optoelectronics Engineering, University of Chinese Academy of Sciences, Beijing 100049, China}

\date{\today}

\begin{abstract}

With the recent report of near ambient superconductivity at room temperature in the N-doped lutetium hydride (Lu-H-N) system, the understanding of cubic Lu-H compounds has attracted worldwide attention. Generally, compared to polycrystal structures with non-negligible impurities, the single-crystalline form of materials with high purity can provide an opportunity to show their hidden properties. However, the experimental synthesis of single-crystalline cubic Lu-H compounds has not been reported thus far. Here, we developed an easy way to synthesize highly pure LuH$_{2+x}$ single-crystalline films by the post-annealing of Lu single-crystalline films (purity of 99.99\%) in the H$_2$ atmosphere. The crystal and electronic structures of films were characterized by X-ray diffraction, Raman spectroscopy, and electrical transport. Interestingly, Lu films are silver-white and metallic, whereas their transformed LuH$_{2+x}$ films become purple-red and insulating, indicating the formation of an unreported electronic state of Lu-H compounds possibly. Our work provides a novel route to synthesize and explore more single-crystalline Lu-H compounds.

\end{abstract}

\maketitle
\newpage

\section{Introduction}

A very recent report of near ambient superconductivity at room temperature in the Lu-H-N system immediately ignited the worldwide interest of studying Lu-H compounds \cite{Nature-2023-Dias,arXiv-2023-Jin,arXiv-2023-LiuP, CPL-2023-Cheng, arXiv-2023-Wen,arXiv-2023-Sun,arXiv-2023-Liu, arXiv-2023-Cheng,arXiv-2023-Wen2,SB-2023-Cheng,arXiv-2023-Boeri,arXiv-2023-Cao,arXiv-2023-MLiu,arXiv-2023-Ho,arXiv-2023-Cui,arXiv-2023-Zurek,arXiv-2023-MM,arXiv-2023-Heil,ArXiv-2020-Duan,Research-2022-Cui,IC-2021-Cui,M-2021-Wang,arXiv-2023-Wen3,arXiv-2023-LuT,arXiv-2023-Errea,arXiv-2023-Fiory,arXiv-2023-Zhang,arXiv-2023-Hemley,arXiv-2023-Monserrat}. Regarding this subject, one fundamental identification is about the composition and structure of the red matter showing near ambient superconductivity at room temperature. In the initial report, the authors assigned the emergent superconductivity to cubic $ Fm\bar{3}m $-LuH$_{3-\delta}$N$_\varepsilon$ (92.25\% purity) with both N-substitution and H-vacancy defects \cite{Nature-2023-Dias}. However, identifying the composition and structure of $ Fm\bar{3}m $-LuH$_{3-\delta}$N$_\varepsilon$ is still a big challenge.

Generally, the stable phase of LuH$_3$ has a rhombohedra structure with space group $P\bar{3}c1$, not the cubic structure with space group $ Fm\bar{3}m $ \cite{arXiv-2023-Ho,arXiv-2023-Boeri}. Therefore, based on a combination of pressure-dependent color, the X-ray diffraction data, and the Raman data, the composition of experimental $Fm\bar{3}m $-LuH$_{3-\delta}$N$_\varepsilon$ was considered to be LuH$_{2\pm{x}}$N$_y$ by several experimental works \cite{arXiv-2023-Wen2,arXiv-2023-Liu,CPL-2023-Cheng}. Moreover, high-throughput first-principles calculations pointed out that there are no thermodynamic stable Lu-H-N ternary structures and the experimental $ Fm\bar{3}m $-LuH$_{3-\delta}$N$_\varepsilon$ may be a mixture of LuH$_2$ and LuN \cite{arXiv-2023-MLiu}. However, a very recent experimental work shows that N-doped $ Fm\bar{3}m $-LuH$_{3}$ can be synthesized by applying a 2 GPa pressure \cite{arXiv-2023-MM}. Furthermore, it is found that both N-doping and pressure can stabilize $ Fm\bar{3}m $-LuH$_{3}$, and Lu$_8$H$_{21}$N is a stable Lu-H-N ternary at 1 GPa \cite{arXiv-2023-Ho}. It is noteworthy that both $ Fm\bar{3}m $-LuH$_{2}$ and $ Fm\bar{3}m $-LuH$_{3}$ are cubic with a lattice parameter $\sim$ 5 \AA ~ \cite{arXiv-2023-Ho,arXiv-2023-Boeri}. Therefore, it is not easy work to distinguish them experimentally. Synthesizing single-crystalline Lu-H compounds is very helpful to understand their intrinsic properties, but it has not been reported thus far.

To address the above, we developed an easy method of synthesizing single-crystalline cubic LuH$_{2+x}$ films by the post-annealing of single-crystalline Lu films in the H$_2$ atmosphere. Unexpectedly, in sharp contrast to metallic LuH$_2$ powders, the LuH$_{2+x}$ films become purple-red and insulating, indicating the formation of an unreported electronic state of Lu-H compounds. Our work provides a novel route to synthesize more single-crystalline Lu-H compounds.

\section{Experiments}

Single-crystalline Lu thin films (thicknesses $\sim$ 80 nm) were grown on (111)-MgO single crystal substrates (5$\times$5$\times$ 0.5 mm$^3$) by a home-made high-pressure radio frequency (RF) magnetron sputtering, the setup of which is analogous to our previous reports \cite{APLM-2021-Zhang,PRM-2021-Bi,ACSP-2021-Zhang,ACSAMI-2021-Zhang}. For this sputtering system, the base vacuum pressure is $\sim$ 1 $\times$10$^{-7}$ Torr, and the purities of the 2-inch Lu target and the Ar reactive gas are 99.99\% and 99.999\%, respectively. During growth, the Ar pressure was kept at 0.02 torr with a gas flow of 5.6 sccm, and the substrate temperature was held at  300 $^\circ$C. The power of the RF generator was 80 W.  Furthermore, to ensure the uniformity of the films, the heating stage was rotated during growth. After deposition, the samples were cooled down to room temperature in the same Ar atmosphere.  Then, the Lu films on MgO substrates were annealed at 280 $^{\circ}$C for 4 hours in a flow of H$_2$/Ar mixture gas (the volume fraction of H$_2$ is 8\%, purity of 99.99\%). During the post-annealing, the pressure of Ar/H$_2$ was kept at 0.14 MPa. To verify the reproducibility of synthesizing single-crystalline LuH$_{2+x}$ films, we repeated  the process of growth and post-annealing more than three times. All films are purple-red, insulating, and stable in the air.

The crystal structures of Lu and LuH$_{2+x}$  films were characterized by a high-resolution X-ray diffraction (XRD, Bruker D8 Discovery) with the Cu K$ _{\alpha} $ source ($ \lambda $ = 1.5406 \AA) and confocal Raman spectroscopy (Renishaw inVia Reflex) using a 532 nm laser. The electrical properties were measured from 300 to 2 K by Physical Property Measurement System (PPMS) in a van der Pauw geometry (DynaCool, Quantum Design).

\section{Results and Discussion}
\begin{figure*}[bhtp]
	\includegraphics[width=0.95\textwidth]{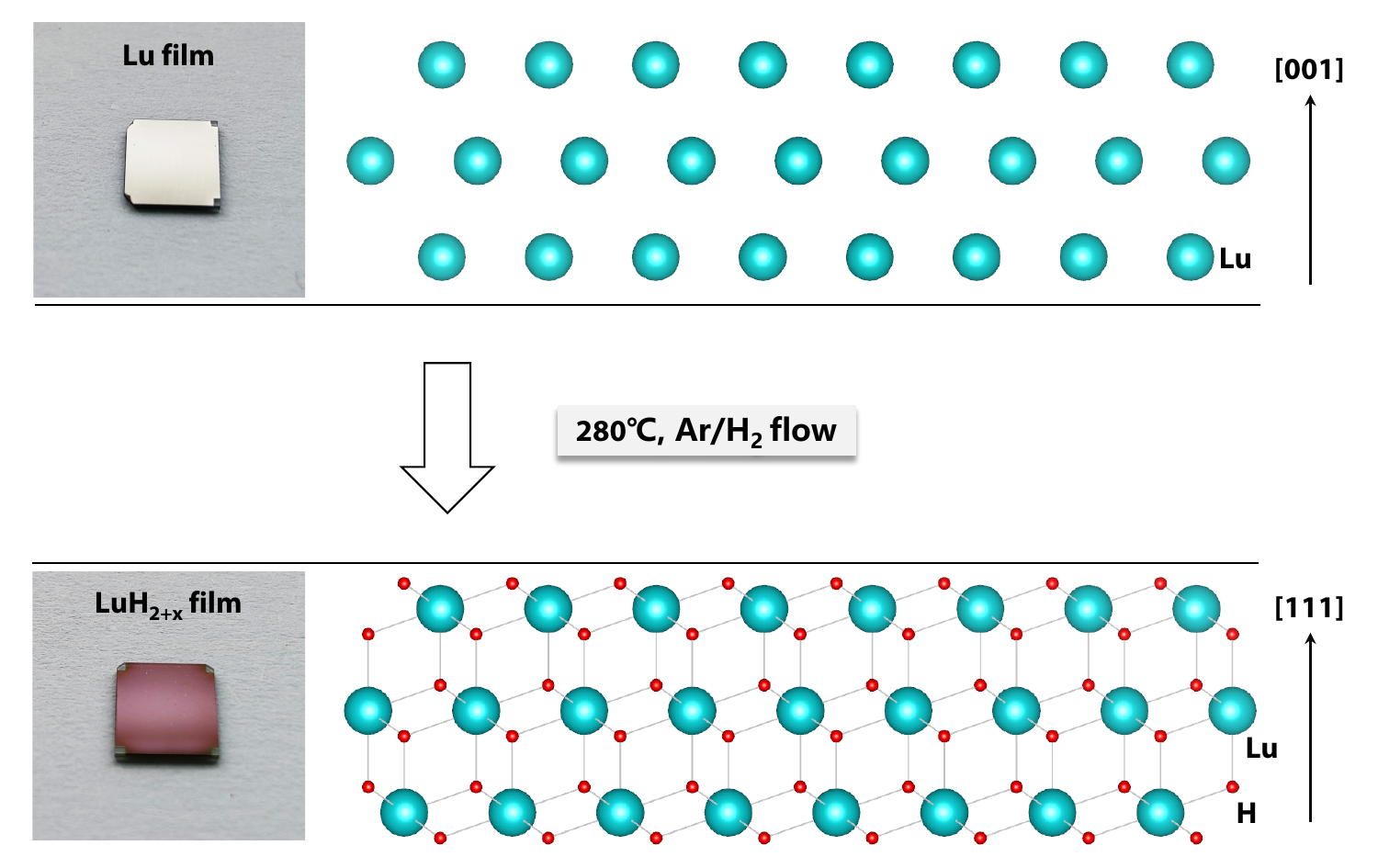}
	\caption{\label{Structure}  The photographs and crystal structures of hexagonal Lu (top panel) and cubic LuH$_{2+x}$ (bottom panel) films.}
\end{figure*}

Figure \ref{Structure} shows the schematic transformation of hexagonal Lu (top panel) to cubic LuH$_{2+x}$ (bottom panel). Generally, single-crystalline Lu has a hexagonal close-packed (hcp) crystal structure with space group $P6_{3}/mmc$. Its lattice parameters are \textit{a} = \textit{b} = 3.516 \AA, and \textit{c} = 5.573 \AA ~\cite{PRB-1998-Vohra}. Experimentally, the color of Lu films is silver-white before post-annealing (see the top panel in Fig. \ref{Structure}). Unexpectedly, arfter post-annealing the color of the films change to purple-red (see the bottom panel in Fig. \ref{Structure}), indicating the formation of Lu-H compounds. However, it has been known that bulk $Fm\bar{3}m$-LuH and $Fm\bar{3}m$-LuH$_3$ are unstable at ambient conditions \cite{arXiv-2023-Boeri}. Therefore, the composition and structure of our purple-red samples are likely to be $Fm\bar{3}m$-LuH$_{2+x}$. Also, it is noted that both $ Fm\bar{3}m $-LuH$_{2}$ and $ Fm\bar{3}m $-LuH$_{3}$ are cubic with a lattice parameter $\sim$ 5 \AA ~\cite{arXiv-2023-Ho,arXiv-2023-Boeri}. Moreover, N-doped $ Fm\bar{3}m $-LuH$_{3}$ has been synthesized at 2 GPa pressure \cite{arXiv-2023-MM}. 

\begin{figure*}[bhtp]
	\includegraphics[width=0.5\textwidth]{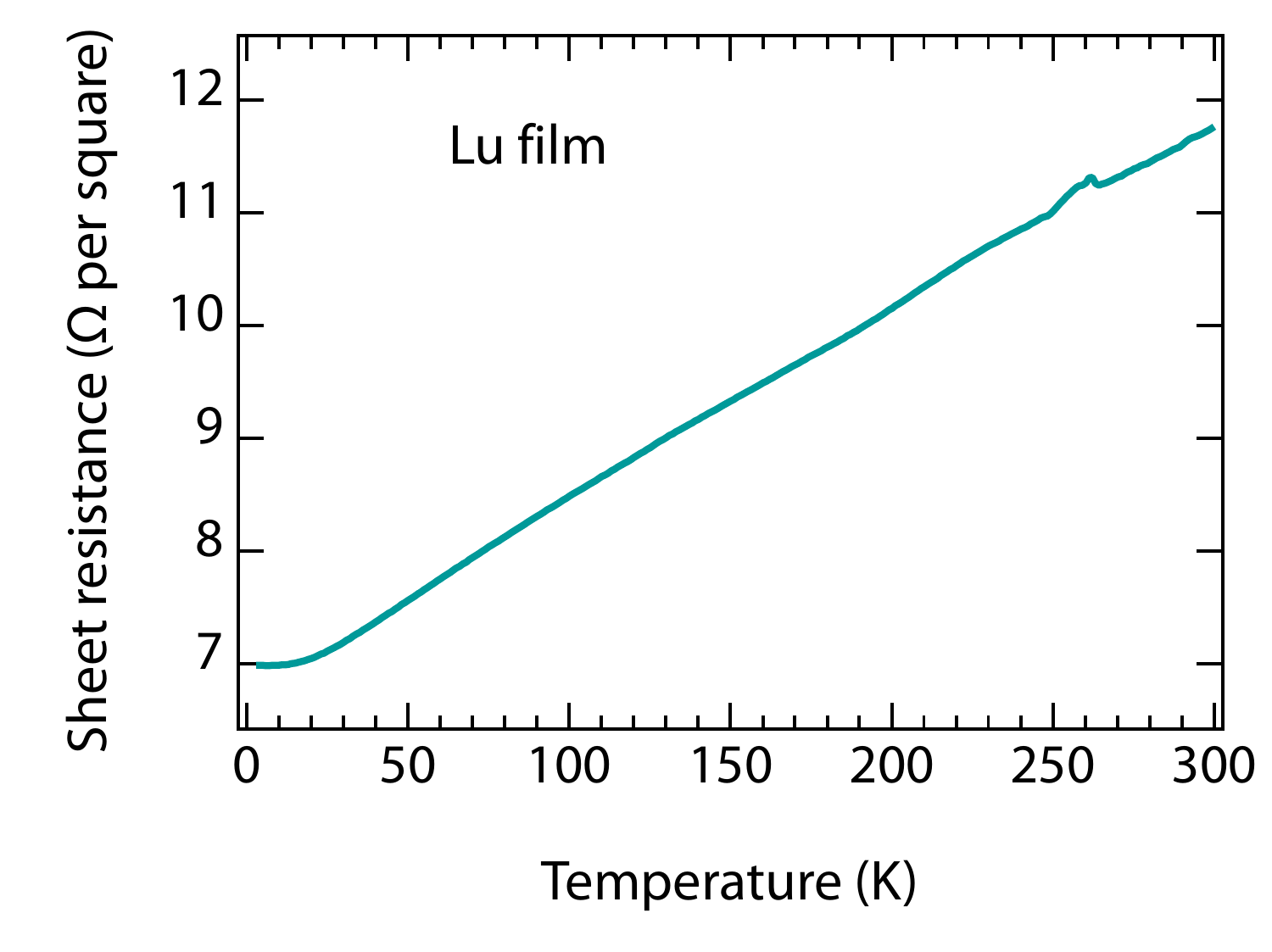}
	\caption{\label{Resistance} Temperature-dependent sheet resistance of the Lu film from 300 to 2 K.}
\end{figure*}

First, to understand the conductivity of Lu and  LuH$_{2+x}$ films, we measured electrical transport from 300 to 2 K. As seen in Fig. \ref{Resistance}, the Lu films are highly conducting, the sheet resistance of which is $\sim$ 11 $\Omega$ at room temperature. With decreasing the temperature from 300 to 2 K, the resistance of Lu films has a linear dependence of the temperature down to 20 K, then becomes a constant below 20 K. This behavior is consistent with our previous observation of polycrystal Lu bulk \cite{arXiv-2023-Cao}. Unexpectedly, in sharp contrast to metalic LuH$_2$ powders and polycrystals \cite{arXiv-2023-Cao,CPL-2023-Cheng, arXiv-2023-Wen,arXiv-2023-Sun,arXiv-2023-Liu, arXiv-2023-Cheng,arXiv-2023-Wen2,SB-2023-Cheng,arXiv-2023-Wen3}, our LuH$_{2+x}$ films are highly insulating with the color of purple-red, indicating the formation of an unreported electronic state of Lu-H compounds. These different properties can result from the differences in hydrogen content in reported LuH$_2$ and our LuH$_{2+x}$. It is noteworthy that the color of LuH$_3$ at 1 bar pressure is red \cite{arXiv-2023-MM}. Therefore, the value of $x$ may be positive for LuH$_{2+x}$.

\begin{figure*}[bhtp]
	\includegraphics[width=0.95\textwidth]{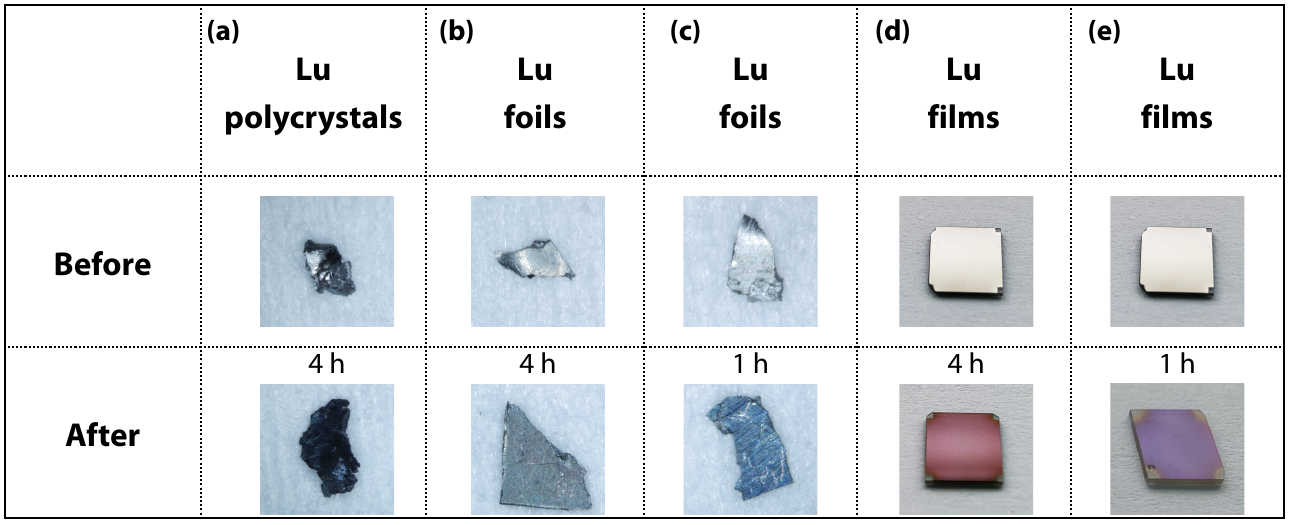}
	\caption{\label{Photograph}  The photographs of Lu (top panel) and LuH$_{2+x}$ (bottom panel) samples. The thickness of Lu foils in (b) and (c) are near 0.1 mm.}
\end{figure*}

To understand the effect of post-annealing on the properties of LuH$_{2+x}$ samples, we prepared the LuH$_{2+x}$ samples by post-annealing Lu polycrystals, Lu foils, and Lu films for 4 and 1 hour. As seen in Fig. \ref{Photograph}, after 4 hours post-annealing, the thick LuH$_{2+x}$ samples prepared from Lu polycrystals and Lu foils are metallic, whereas the LuH$_{2+x}$ films prepared from the Lu films are insulating. With decreasing the annealing times to 1 hour, the thick LuH$_{2+x}$ samples prepared from Lu foils are still metallic but show dark blue color, whereas the LuH$_{2+x}$ films prepared from the Lu films are still insulating but become purple.

\begin{figure*}[bhtp]
	\includegraphics[width=0.95\textwidth]{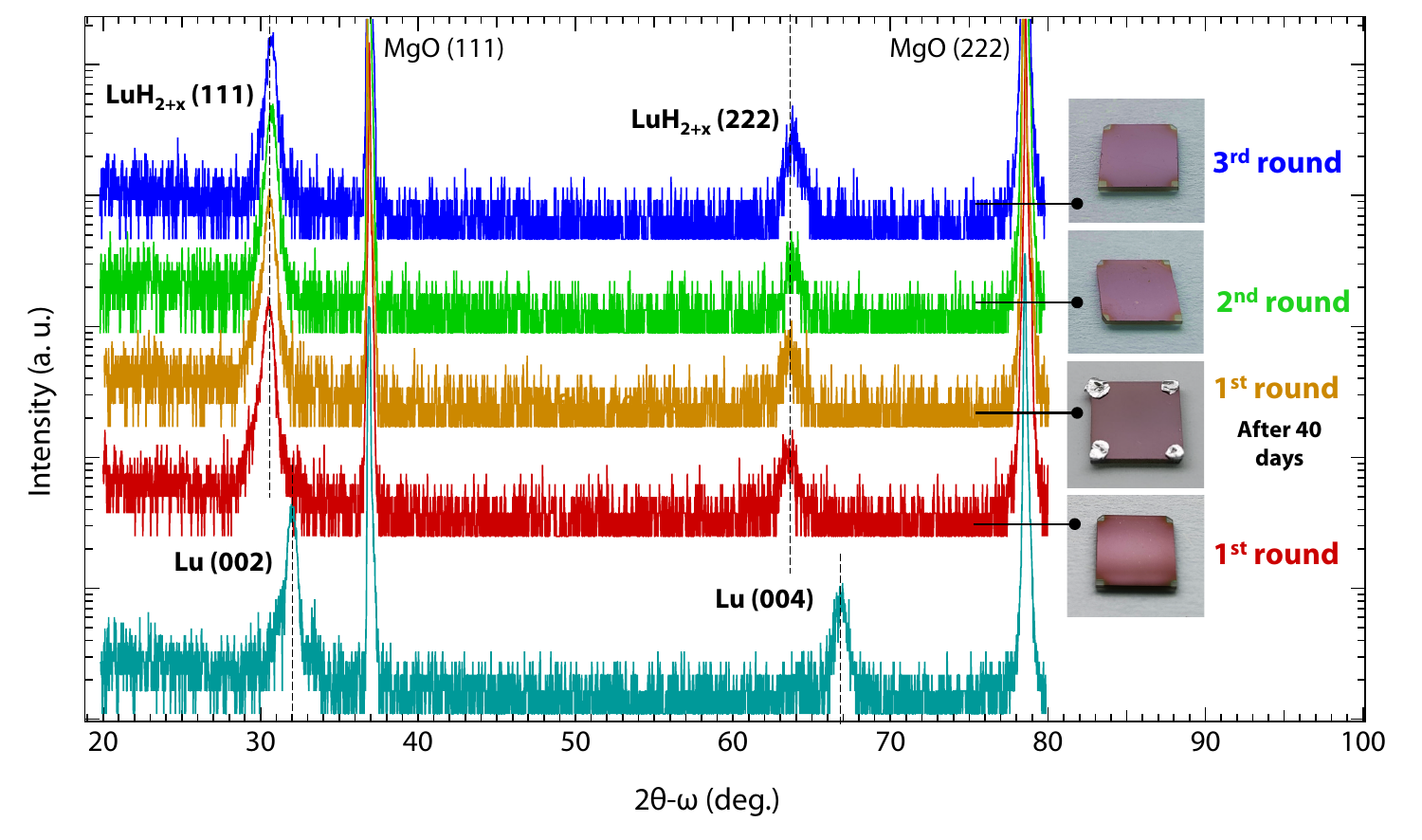}
	\caption{\label{XRD}   Wide-range 2$\theta$-$\omega$ scans of LuH$_{2+x}$ (red, green and bule, three rounds of synthesis; and yellow, after 40 days in air ) and Lu (cyan) films on (111)-MgO substrates, the inserts are the photographs of  LuH$_{2+x}$ films.}
\end{figure*}

Next, to investigate the crystal structures of Lu and LuH$_{2+x}$ films, we carried out XRD characterization. As seen in Fig.\ref{XRD}, the wide-range 2$\theta$-$\omega$ scan on Lu films (see the cyan curve) only shows the (002) and (004) diffraction peaks without detectable secondary phases, indicating the epitaxial growth of single-crystalline Lu films. The peak position of (002) peak is around 31.998$^{\circ}$, corresponding to the lattice parameter $d_{002}$ = 2.495 \AA, which is close to the value of bulk Lu ($d_{002}$ = 2.491 \AA), indicating the epitaxy of Lu films on MgO substrates. Interestingly, after the post-annealing, the two diffraction peaks have significant shifts to low diffraction angles (see the red curve in Fig. \ref{XRD}). The peak positions of LuH$_{2+x}$ are around 30.478$^{\circ}$ and 63.398$^{\circ}$,  respectively, which agree well with the (111) and (222) diffractions peak positions of cubic LuH$_2$ or cubic LuH$_3$. The extracted out-of-plane lattice parameter of our LuH$_{2+x}$ films is around 5.07 Å, which is consistent to the value of bulk LuH$_2$ or LuH$_3$ ($\sim$ 5 \AA) ~\cite{arXiv-2023-Ho,arXiv-2023-Boeri}. Moreover, to verify the reproducibility of synthesizing single-crystalline LuH$_{2+x}$ films, we repeated the process of growth and post-annealing more than three times, the XRD features of these three samples nearly the same (see the red, green and bule curve in Fig. \ref{XRD}). All films are purple-red, insulating, and stable in the air (see the yellow curve and the inserts in Fig. \ref{XRD}).

\begin{figure*}[bhtp]
	\includegraphics[width=0.95\textwidth]{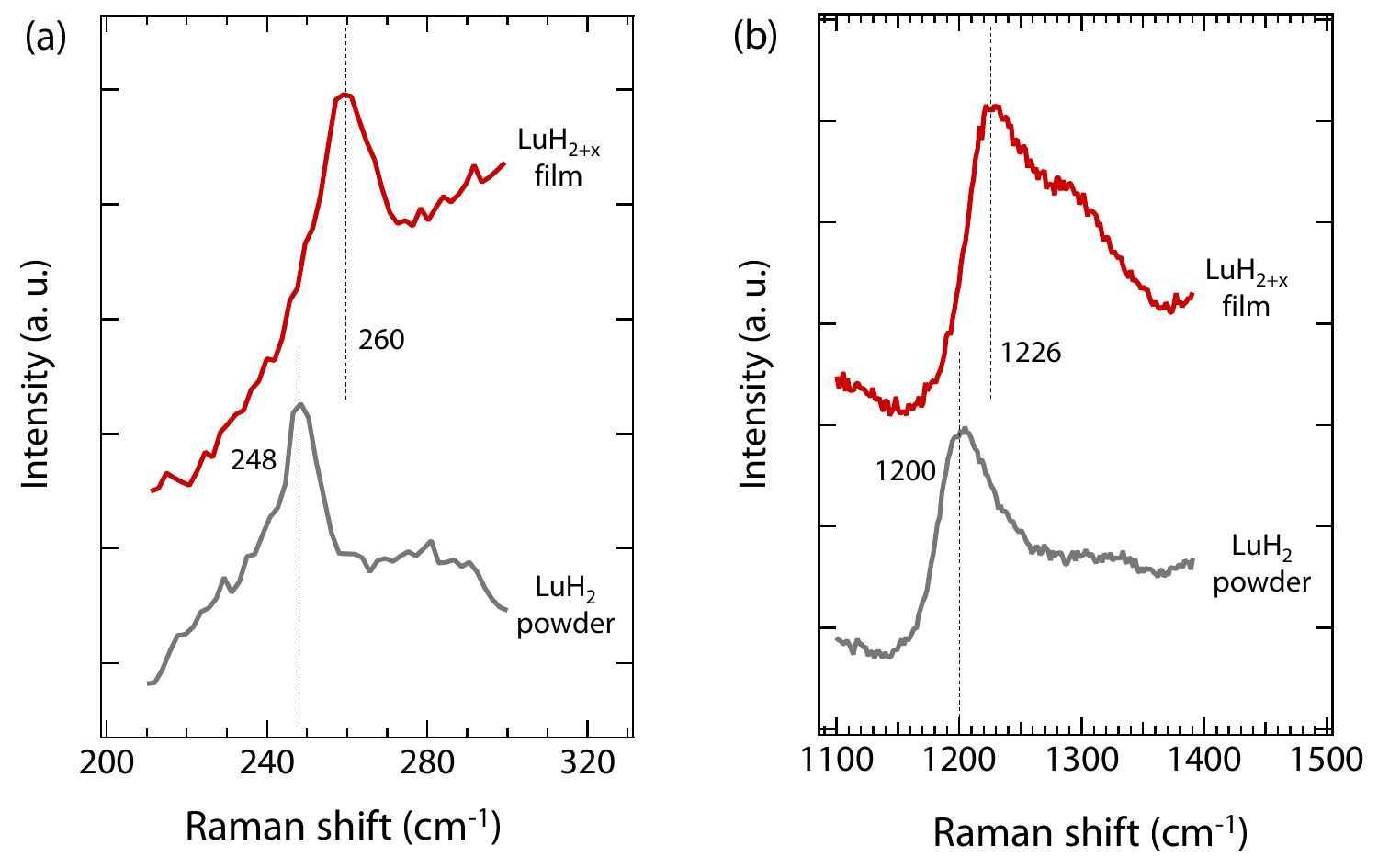}
	\caption{\label{Raman} Raman spectra of LuH$_{2+x}$ films (red curves) and LuH$_{2}$ powders (grey curves).}
\end{figure*}

To further understand the crystal structure of LuH$_{2+x}$ films, we measured Raman spectroscopy on both LuH$_{2+x}$ films and LuH$_2$ powders. It is noted that the contribution of MgO single-crystalline is as small as being ignored. Therefore, the Raman signals of LuH$_{2+x}$ films are strong. As seen in Fig. \ref{Raman}, the two characteristic peaks ($\sim$ 250 and 1200 cm$^{-1}$) can be observed in both LuH$_{2+x}$ films and LuH$_2$ powders. In contrast to the reported peak positions (248 and 1200 cm$^{-1}$) of LuH$_2$ powders and polycrystals \cite{Nature-2023-Dias,CPL-2023-Cheng,arXiv-2023-Liu,arXiv-2023-MM}, these two Raman characteristic  peaks have a blue shift to $\sim$ 260 and 1226 cm$^{-1}$, respectively. It has been pointed out that the strong mode 1200 cm$^{-1}$ can be associated with the T$_{2g}$ Raman-active mode of the cubic structure \cite{arXiv-2023-MM}, further demonstrating the formation of cubic LuH$_{2+x}$. Based on the purple-red color, the insulating state, and the cubic structure of our LuH$_{2+x}$ films, it indicates a formation of an unreported electronic state in LuH$_{2+x}$ films.

\section{Conclusion}

In this work, we developed an easy way to synthesize single-crystalline cubic LuH$_{2+x}$ films by the post-annealing of single-crystalline Lu films in the H$_2$ atmosphere. The crystal and electronic structures of films were characterized by XRD, Raman, and electrical transport. Interestingly, Lu films are silver-white and metallic, whereas their transformed LuH$_{2+x}$ films become purple-red and insulating. Based on the purple-red color, the insulating state, and the cubic structure of LuH$_{2+x}$ films, it indicates the formation of an unreported electronic state in our LuH$_{2+x}$ films. Our work provides a novel route to synthesize and explore more single-crystalline Lu-H compounds.

\section{Acknowledgments}

We acknowledge insightful discussions with Rui Peng. This work was supported by the National Key R\&D Program of China (Grant No. 2022YFA1403000), the National Natural Science Foundation of China (Grant Nos. U2032126 and 11874058), the Pioneer Hundred Talents Program of the Chinese Academy of Sciences, the Zhejiang Provincial Natural Science Foundation of China under Grant No. LXR22E020001, the Beijing National Laboratory for Condensed Matter Physics, the Ningbo Natural Science Foundation (Grant No. 20221JCGY010338), and the Ningbo Science and Technology Bureau (Grant No. 2022Z086).

\newpage

\end{document}